# Electrosense: Open and Big Spectrum Data

Sreeraj Rajendran*‡‡, Roberto Calvo-Palomino‡‡¶§, Markus Fuchs‡, Bertold Van den Bergh*,
Héctor Cordobés¶, Domenico Giustiniano¶, Sofie Pollin* and Vincent Lenders∥
Email: {bertold.vandenbergh, sreeraj.rajendran, sofie.pollin}@esat.kuleuven.be
{domenico.giustiniano, hector.cordobes, roberto.calvo}@imdea.org
fuchs@sero-systems.de, vincent.lenders@armasuisse.ch.
*Department ESAT, KU Leuven, Belgium, ¶IMDEA Networks Institute, Madrid, Spain,
‡SeRo Systems, Germany, §Universidad Carlos III of Madrid, Spain, ∥armasuisse, Thun, Switzerland.

*Abstract*—While the radio spectrum allocation is well regulated, there is little knowledge about its actual utilization over time and space. This limitation hinders taking effective actions in various applications including cognitive radios, electrosmog monitoring, and law enforcement. We introduce Electrosense, an initiative that seeks a more efficient, safe and reliable monitoring of the electromagnetic space by improving the accessibility of spectrum data for the general public. A collaborative spectrum monitoring network is designed that monitors the spectrum at large scale with low-cost spectrum sensing nodes. The large set of data is stored and processed in a big data architecture and provided back to the community with an *open spectrum data as a service* model, that allows users to build diverse and novel applications with different requirements. We illustrate useful usage scenarios of the Electrosense data.

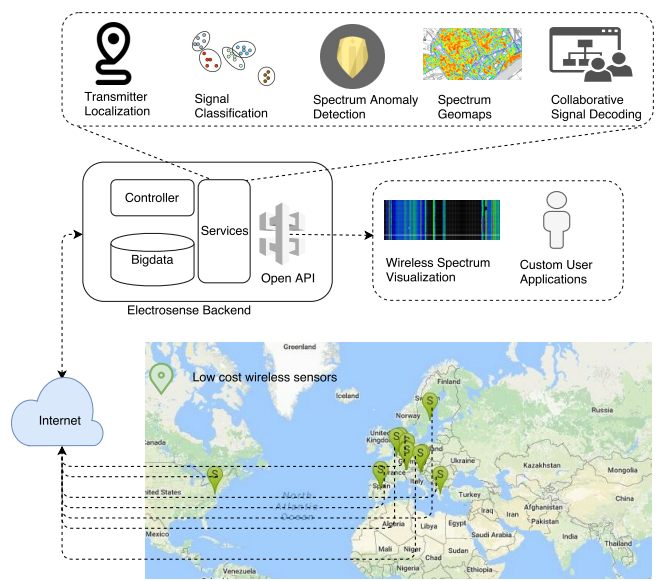

Fig. 1: High-level overview of the Electrosense network: Low-cost sensors collect spectrum information which are sent to the Electrosense backend. Different algorithms are run on the collected information in the backend and the results of these algorithms are provided to the users as a service through an open API. Users can develop their own applications from the spectrum information retrieved using the API.

## I. INTRODUCTION

Over the past years, we have seen a tremendous increase in mobile data usage. To meet the data demands, more radio frequency (RF) bands are being used, cells become smaller, data rates increase and more users accommodated. In addition, novel technologies are being proposed that can coexist with legacy technology and modulation types, and short and long range Internet-of-Things (IoT) communications increase the variety of physical and medium access protocols connected to the Internet. As a result, RF spectrum use is fragmented, bursty and very diverse. The monitoring of this spectrum is a complex task, as it requires a very dense sampling in time, frequency and space, resulting in a radio spectrum data deluge. Nevertheless, with the increasing spectrum usage complexity, knowledge and understanding of the spectrum usage patterns is becoming more and more critical to ensure continued effective use of this scarce resource.

Spectrum resource monitoring is important for end users, operators, spectrum regulatory bodies and also military applications. Each use case has its own specific needs and challenges. The grand challenge is how to design a cost-effective solution that meets the requirements of all potential end users. Users might be interested in electrosmog or optimization of their indoor WiFi network. Regulatory bodies might be keen on enforcing spectrum regulation. Operators might be concerned about coverage maps over time for optimizing their cell networks or refarming of their frequency bands. Military applications might be the most challenging, requiring the detection and positioning of any signal hidden on purpose. Novel operators might be interested in IoT cases, such as cooperative detection of signals transmitted by low-cost low-power transceivers. This paper looks at one key aspect: is it possible to design a spectrum sensing network that meets all the aforementioned requirements and can be deployed quickly by synergistic cooperation between all stakeholders?

Even though a majority of wireless researchers, industries and spectral regulators are keen to develop a worldwide spectrum monitoring infrastructure, albeit several attempts, the research community has not succeeded in deploying one. The multidisciplinary nature of the spectrum monitoring solution is one of the main challenges that prevents the realization of such a system, which in turn requires proper integration of new disruptive technologies. The infrastructure should flexibly address the variability and cost of the used sensors, large spectrum data management, sensor reliability, security and privacy concerns, which can also target a wide variety of the use cases as mentioned before.

‡‡Equal contributors

A few spectrum monitoring solutions are proposed in the literature. Some examples include, Microsoft Spectrum Observatory[1] that allows to sense the spectrum using expensive sensors, Google spectrum[2] for measurements on TV whitespaces and the IBM Horizon[3] project that proposed a generic decentralized architecture to share *IoT* data. While Google Spectrum and Blue Horizon fail to cover a large part of the spectrum as they are application specific deployments, Microsoft observatory on the other hand fails to enable a large scale sensor deployment mainly due to the cost of the sensing stations.

We introduce and design Electrosense[4]. Electrosense follows the crowdsourcing paradigm to collect and analyze spectrum data using low-cost sensors as main device for sensing the spectrum. The main goal of this initiative, as shown in Figure 1, is to sense the whole spectrum in different regions of the world and provide the processed spectrum data to any user interested to acquire a deeper knowledge of the spectrum usage, enabling applications as the ones mentioned above [1,2]. Worldwide deployments are plausible when the crowdsourcing paradigm is combined with low cost sensors. Electrosense sensors are designed using inexpensive and easily accessible software-defined radio (SDR) front-ends and embedded platforms like Raspberry Pi, to reduce the cost that spectrum data contributors have to bear. A low-cost down-converter design[5] is also provided to extend the range of the low-end SDR platforms from 1.7 GHz to 6 GHz. Even though the aim of the Electrosense project is to use low-cost hardware, high-end SDR devices can also be part of the network. Open source software modules for low- and high-end sensors are provided by Electrosense for easy setup[6].

Crowdsourcing initiatives have been successfully applied in other sensing contexts, for example, distributed sensor networks collecting information about temperature, air quality, pollution or air traffic are now in widespread use today. Yet, the radio spectrum data collection poses novel challenges because of the sheer volume of information in the spectrum, which is several orders of magnitude higher than the data collected in typical sensing contexts as mentioned above. The amount of data produced by a single sensor typically varies between 50 Kb/s and 50 Mb/s depending on the sensor platform and configuration. Storing spectrum data received from each sensor can swiftly get any server machine out of storage space, for instance data from 60 sensors for a single month needs a storage space of 1 terabyte in 50 Kb/s mode. In order to handle this large volume of data in the backend and extract meaningful information over the entire spectrum, a flexible big data architecture is designed. This spectrum management architecture is responsible for the sensors' control, data storage, and algorithm deployment in the back-end for further processing.

Electrosense also includes a complete framework for identifying, locating and deploying sensors securely through a consistent registration process and remote control framework of the sensors. The sensed spectrum data from different sensors can be retrieved from the electrosense backend using an open API[7] over the Internet. Data privacy concerns are addressed by enabling public, private and restricted data access permissions which restrict user data access with suitable time resolutions. Readily available spectrum aggregation tools in the back-end help the network users, to do easy spectrum data analysis as an additional incentive. The data access permissions and other applications are detailed in Section II and Section IV of this paper.

The rest of the paper is organized as follows. A brief overview of the service model and the open API are presented in Section II. Section III defines Electrosense system design considerations and the final architecture. Section IV details the spectrum data analysis tools and applications. Finally, conclusions and future work are presented in Section V.

## II. SPECTRUM DATA AS A SERVICE

### A. Service Model

In order to make the most out of the data, we introduce an Open Spectrum Data as a Service (OSDaaS) model in which the spectrum data can be used by several applications, each of them with unique requirements. This approach differs from a classical "Infrastructure as a Service" model used in other contexts where different applications cannot run at the same time due to conflicting requirements. The spectrum data inherently changes over time and thus allocating the node to a specific application would reduce the information and the interest of several users interested in the same spectrum data. Therefore, the current framework prevents the users from launching their own measurement campaigns as they only need to access gathered data from one of the two pipelines of the sensor: PSD (Power Spectral Density) and IQ (in-phase and quadrature components of raw signals). In contrast, the flexibility of the OSDaaS model is an incentive to join and use the network and an opportunity for a wide engagement and participation of citizens. In addition, the spectrum data gathered in the past can be re-processed with a new application in mind or new spectrum data can be compared with older spectrum data. Users can join the Electrosense network easily by adding their own sensor through the web interface[8]. Specifics about the sensor such as antenna details and sensor location can be specified during the registration process. The sensor status and location are readily visible in a global sensor map once it is registered.

### B. Application Programming Interface (API)

Alongside with data processing, it is important to provide the community with easy access to the data. An open API

---

[1] http://observatory.microsoftspectrum.com/
[2] https://www.google.com/get/spectrumdatabase/
[3] https://bluehorizon.network/documentation/sdr-radio-spectrum-analysis
[4] https://electrosense.org/
[5] https://github.com/electrosense/hardware
[6] https://github.com/electrosense
[7] https://electrosense.org/open-api-spec.html
[8] https://electrosense.org/join.html

serves this purpose for bulk or streaming data retrieval. The API also allows access to the algorithms' output running in the backend, for example results of the modulation classification algorithms or the anomaly detector. Electrosense users can retrieve magnitude data along with the sensor details using the API. The API allows two data retrieval modes: aggregated and raw Fast Fourier Transform (FFT) data. Aggregated query type allows the user to request for a bulk of data with specified frequency and time resolutions, after applying a predefined aggregation function such as averaging or max-value. Raw requests permit to retrieve magnitude FFT data acquired by a sensor as such without any modifications. Raw FFT data from a sensor can be accessed only by its owner. In addition, as shown in Figure 3, there is an IQ data pipeline which is enabled on demand for data testing or for algorithms that work on IQ data in the backend. The IQ data from a sensor is only directly available to its owner through the API, but conclusions from algorithms related to IQ data will be made available to all users. Furthermore, real-time spectrum monitoring is possible through a streaming API in the backend. A few examples for data retrieval can be found in the repository[9].

### C. Security and Privacy

Wideband spectrum monitoring on a large scale typically raises critical security and privacy concerns. Security concerns include secure data transmissions from the sensor and proper sensor identification. Data privacy concerns in terms of data access restrictions should also be addressed. In addition, capture and storage of IQ data, which can be decoded to reveal content, is not advisable especially in the military bands. Electrosense's design addresses sensor identification via a proper registration process. The data from Electrosense sensors are sent over secure TLS channels which guarantees data privacy and integrity and prevents the data getting modified by an attacker. The IQ pipeline is only accessible by the Electrosense backend and the data is deleted once the backend algorithms are done processing it. In addition, IQ data is only made available to the users for their own sensors through the API. Furthermore, the highest time and frequency resolution that users can get from other sensors over the API are limited to 60 seconds and 100 kHz respectively. Electrosense also allows to obfuscate sensor's publicly displayed location within a range of a few kilometers in the sensor map, thus protecting users' locational privacy.

## III. SYSTEM ARCHITECTURE

Electrosense was designed with a clear vision in mind that translates into a list of design goals as described below. We present the actual architecture chosen to meet these requirements and give a more detailed explanation of its components.

### A. Design Goals and Electrosense Vision

Electrosense is a crowdsourcing approach towards spectrum monitoring where volunteers play a central role. To keep the entry barrier as low as possible, large-scale deployment can only be achieved with *low-cost sensors*. The backend should be flexible, to enable *wide-band monitoring* from sweeping low-end sensors, to *real-time scanning* with more capable SDR platforms or spectrum analyzers. In addition, the backend should be horizontally *scalable*, enabling to grow the network continuously. Scalability introduces even more complexity into the system, which makes it prone to failures. As such, the architecture must also be *fault-tolerant*. Finally, the backend should be able to process a large amount of data from any sensor with *low latency*.

[9]https://github.com/electrosense/api-examples

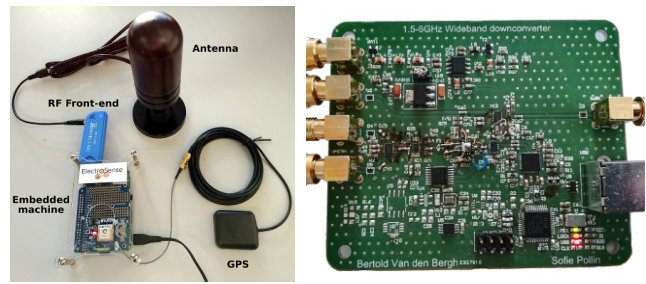

Fig. 2: [Left] Sensor based on a single-board computer (RPi-2) and RF front-end (RTL-SDR) device responsible for signal acquisition. [Right] First prototype of custom RF converter to enable spectrum scanning in 0-6 GHz range.

With its main goal to serve researchers as a platform for spectrum analysis, the system should have a *central component to control* the sensors for specific measurement campaigns. Some applications might need to give immediate response while others generate insights by inspecting the data set as a whole. Following these requirements, the system should support *low-latency stream processing* and *large-scale batch analyses* at the same time.

In order to address these goals and requirements, the Electrosense architecture consists of three main components: (i) the sensors deployed by users collecting spectrum measurements, (ii) a centralized controller infrastructure to interact with the sensors and to administrate measurement campaigns and (iii) the backend which is responsible for collecting data from all the sensors and provide insights by applying algorithms. The following subsections describe these components in detail.

### B. Sensor

The sensing nodes used in the Electrosense network consist of small-sized, low-cost, software-defined embedded computing devices connected to a simplistic RF front-end and a general purpose antenna [3] as shown in Figure 2 (left). The sensors can measure the spectrum ranging from 20 MHz up to 6 GHz using an optional down-converter (Figure 2 right). The sensors include an optional low-cost GPS device to synchronize the time among them which in turn helps in enabling collaborative spectrum scanning and detection algorithms. For instance, in [1] the feasibility to recombine signals collaboratively from different sensors, using GPS as a reference clock, is analyzed in detail.

Two signal pre-processing pipelines are enabled on the sensor as illustrated in Figure 3. Each sensor can be configured in order to work with both PSD or IQ pipeline. When the sensor is configured in the PSD mode, the spectrum sensed by the RF front-end is converted to the frequency domain using a Fast Fourier Transform (FFT). In this mode only the averaged squared magnitude FFTs are sent to the backend. Depending on the averaging and compression factors, the produced data is in the order of 50-100 Kb/s. This pipeline saves bandwidth and storage over the IQ pipeline and thus contributes to the cost-efficiency of the whole network. In addition, the on-board GPU on the Raspberry Pi is used for computing FFTs, reducing the processing time. The PSD mode aids spectrum data retrieval using a restricted network bandwidth which enables applications where phase information is not crucial to analyze data in terms of signal power (Section IV-B). In addition, wideband scanning techniques are applied that prioritize frequency bands of bursty activities over bands which are stationary [3] to overcome the hardware-limitations of low-cost radios, for instance the limited sampling bandwidth of the low-end ADCs (e.g 2.4 MHz for RTL-SDR).

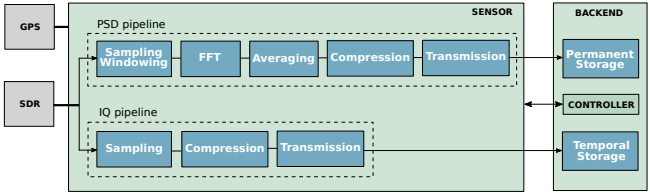

Fig. 3: Sensor architecture: Sensor software contains two different pipelines for retrieving spectrum information.

For applications which require phase and non-averaged information, the sensor implements an IQ pipeline. The IQ mode retrieves raw measurements (IQ samples) from the RF front-end that are then compressed and sent to the backend. This pipeline easily produces up to 50 Mb/s and hence the data is only stored temporarily in the backend. Besides spectrum measurements in both pipelines, the sensor also provides timing information together with RF-specific values such as antenna gain. The data structure used to transport results over the Internet is flexible and can be extended in the future.

The software that runs in the electrosense nodes is released as open source[10]. In addition to this, a GNU Radio based software module, gr-electrosense[11], is provided to allow high-end SDR users to be a part of the Electrosense network.

### C. Controller

Apart from continuous wide-band monitoring, it is interesting to have a more detailed look at particular parts of the frequency spectrum. This is accomplished with the help of a command-and-control layer as depicted in Figure 4. In order to start and control such in-depth measurements, the controller communicates with sensors directly and influences their scanning strategy, some examples include, the frequency range the sensor should scan, the sensor frequency hopping strategy or the sensor sampling rate.

The core of the controller infrastructure is a Message Queue Telemetry Transport (MQTT) broker system which consists of a set of several independent machines forming a cluster. This ensures scalability and fault-tolerance in case of a hardware failure. MQTT is a publish-subscribe-based messaging protocol and has a wide adoption in IoT applications [4]. The brokers have a permanent connection to the sensors over secure TLS channels as well as to a machine called *master controller*. The master offers an interface where administrators can start and stop *measurement campaigns*. It is responsible for executing these commands on different sensors using the MQTT connection. Spectrum measurements collected by the sensors are passed to the backend over secure TLS channels as shown in Figure 4. The control layer is a part of the Electrosense backend and works in a closed loop with the data processing path for easy response validation.

[10]https://github.com/electrosense/es-sensor
[11]https://github.com/electrosense/gr-electrosensexamplese

### D. Backend

All data collected by the sensors is sent to the *backend*. In order to fulfill the design goals, in particular scalability, the system is built upon a Lambda architecture [5] with exclusively horizontally scalable system components. The Lambda architecture is a three-layered architecture consisting of *batch*, *speed* and *serving* layer. An overview of the whole backend architecture can be found in Figure 4. The batch layer is responsible for executing bulk analysis tasks on large data sets. These tasks, mostly involve complex long-running algorithms, whose results might take several minutes to hours to complete. To overcome this gap and, furthermore, provide the capability for real-time monitoring with sub-second delay, there is the need for a speed layer. It is possible to run different or identical applications on each of these layers simultaneously.

In addition, the Electrosense backend has an *ingestion* layer to support high availability. It consists of a distributed message queue that stores multiple replicas of each incoming measurement. Implementation and functionality of each layer is detailed in the subsequent sections.

*1) Ingestion Layer:* Sensor measurements eventually reach the *collector* as shown in Figure 4. The collector inserts received data as messages into a distributed queuing system. Using a distributed queuing system, we decouple data ingestion from processing in the batch and speed layer, allowing an asynchronous mode of operation. Furthermore, the ingestion layer serves as a intermediate buffer for incoming data, allowing maintenance of batch and speed layers without data loss. Apache Kafka [6], a distributed messaging system which features replication, partitioning and data retention is used as the queuing system. Following the ingestion layer, spectrum data is processed on two different paths: in batch and speed layer simultaneously.

*2) Batch Layer:* The batch layer is responsible for storing all incoming raw data as an immutable master dataset and executing long-running workloads. The batch layer facilitates processing of huge amounts of historical data, ensuring no loss of information in case of algorithmic errors. The master dataset is stored in a binary format on a distributed file system. The Hadoop Distributed File System (HDFS) [7] is employed allowing data to be stored in files which are distributed and replicated over a cluster of servers. Apache Spark [8], which allows parallel computation by evenly distributing workload over several computing nodes, is employed for data processing. For accessing data stored on HDFS, a fast and scalable query engine Cloudera Impala [9] which understands a subset of the SQL language is deployed. The latency of the batch layer as a whole is in the order of minutes or even hours.

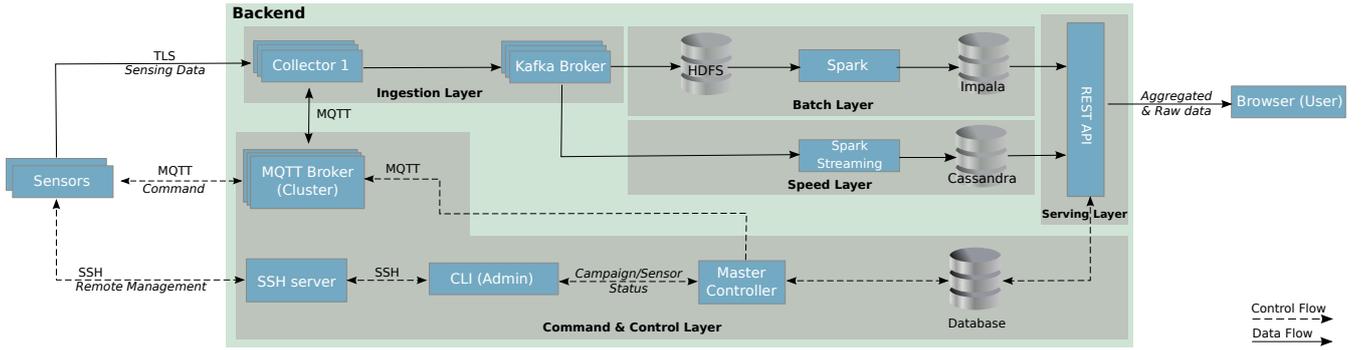

Fig. 4: Sensors are managed by the control layer which sets the scanning parameters for a specific campaign. The configured sensors send the spectrum information to the backend, where the data is processed and distributed over the API to the users.

*3) Speed Layer:* The Electrosense speed layer is based on Spark Streaming. It is an extension of Spark which computes its' results continuously on a parallel cluster over a small window of recent data, with the current default window length set to 5 seconds. Spark Streaming meets the design requirements as the spectrum data is high in volume and algorithms do not demand sub-second latency. As Impala is not suitable for near real-time storage requirements, the speed layer persists its results in another database. We use Apache Cassandra [10], a hybrid between a key-value and column-oriented distributed storage system for this real-time database as it represents well the data structure of spectrum data.

*4) Serving Layer:* As seen before, the batch and speed layers store their results differently. The serving layer takes care of data fusion to provide query results to the user while hiding the complexity of the multi-layer architecture. It offers the open API which serves as the endpoint for any user or application to retrieve data. The serving layer handles queries on different views and combines results from batch and speed layer if necessary. In case results are available in both layers, batch layer results are preferred as they provide more accurate data. In Electrosense, the serving layer is a custom component which offers a RESTful web service over HTTP.

## IV. SPECTRUM DATA PROCESSING AND ANALYSIS

To substantiate the value of the data generated by the Electrosense framework, a few applications are presented in this section. We envision some of these applications to be integrated in Electrosense's batch and speed layer in the backend, but the goal of Electrosense is that users can start implementing their own applications by using the open API.

### A. Live and Historical Spectrum Visualization

We have developed a web front-end to enable easy visualization of the spectrum data[12]. The spectrum visualizer allows interactive sensor selection and varying frequency resolution for detailed spectrum viewing. The visualization tool allows to look at historical spectrum data from the sensors, since the sensor registration. For this purpose, the batch layer pre-computes spectrum data at various temporal and frequency aggregation levels and stores this information in queryable tables for low-latency access over the web. For live spectrum updates with 5 second delays, a streaming display is incorporated to the web interface directly from the speed layer.

[12]https://electrosense.org/app.html

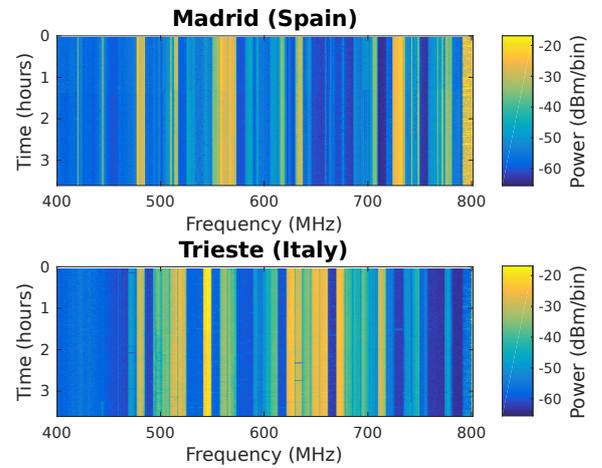

Fig. 5: TV band occupancy in Madrid and Trieste.

### B. Dynamic Spectrum Access

Dynamic Spectrum Access (DSA) enables a technology model to manage spectrum availability and needs in terms of frequency, time, geography, quality and cost of the service. Cooperative and distributed spectrum sensing approaches for cognitive radios can solve hidden primary user problems and reduce the non-detection probability and false alarms remarkably [11]. Advanced collaborative sensing strategies along with detailed sensor density analysis is required to address the complete cognitive cycle. A white space detection study is done to validate the feasibility of the framework in this direction.

The term *white spaces* usually refers to the unused broadcasting frequencies in the TV band (400-800 MHz). The availability of white spaces is highly region dependent, for instance the white spaces in Europe in the 470-790 MHz band is found to be less than in the USA [12]. Using the Electrosense infrastructure and the sensors configured in the PSD pipeline, it is feasible to detect the white spaces in a certain region. Figure 5 shows the spectrum occupancy in the broadcasting TV band (400-800 MHz) of two different cities in Europe with a frequency resolution of 1 MHz and 60 seconds time resolution. Similar analysis can be easily performed by retrieving the aggregated data from the Electrosense backend through the API.

## C. Spectrum Cop

Spectrum enforcement is challenging mainly due to the effort involved in analyzing, detecting and locating anomalies occurring in the spectrum. Even with a large scale deployment of spectrum scanners, the anomaly detection process is demanding. Automated systems to detect pirate FM stations, fake GSM towers, unauthorized transmissions hindering normal functioning of meteorological radars and even transmissions at higher power levels than the permitted levels of operation is fundamental for effective enforcement. A good anomaly detector should first detect whether a particular transmission type is allowed in the frequency of interest. Moreover, it should fingerprint normal spectrum transmissions by analyzing the temporal and spatial features of the transmitters in terms of power levels and spectral occupancy. The former can be achieved to some extent using a modulation classifier and the later by learning spectral occupancy distributions over time. A drastic variation of any of the aforementioned features can be classified as an anomaly.

To validate modulation classification using Electrosense, a time domain deep learning modulation classifier is used in the backend which works effectively with gr-electrosense using the IQ pipeline [13,14]. The deep learning model takes IQ samples as input giving out the probability of the data belonging to a particular modulation class. The model is trained to learn from the modulation schemes' time domain amplitude and phase information, without requiring expert features, such as higher order cyclic moments. Analyses show that the proposed model yields an average classification accuracy of over 90% at varying SNR conditions ranging from 0 dB to 20 dB, independent of channel characteristics. In future a computationally efficient binarized version of the model will be deployed on low-cost sensors reducing the data overhead of sending IQ samples to the backend [14].

## D. Localization

Transmitter localization is a promising application used for generating both automated transmitter maps and transmitter fingerprinting. Research on received signal strength indicator (RSSI)-based indoor and outdoor localization has gained interest in recent years [15]. However, the accuracy of the RSSI information drawn from the Electrosense sensors must be verified to validate localization. Such a sensor is calibrated and then the system gain, consisting of the antenna gain, cable losses and front-end gain, is measured for absolute power measurements. RSSI measurements are made in the 435 MHz ISM band using standard LoRa sensors. A splitter is employed to ensure an equal antenna signal reaches both a LoRa receiver and a low-cost Electrosense sensor. A mobile LoRa transmitter is configured to send packets every three second over a period of 80 minutes. Their GPS logged and time-stamped locations are sent to the receiver. The collected RSS measurements from both the LoRa and Electrosense sensors are depicted in Figure 6. This plot shows that accurate RSSI information can be obtained from properly calibrated Electrosense sensors. The RSSI dips in the sensor data are attributed to the lengthy frequency scan of the sensor, which is over 40 seconds, causing it to miss some packets. Existing algorithms presented in the literature [15] can be deployed using the sensed spectrum data retrieved through the open API or implemented directly in the backend.

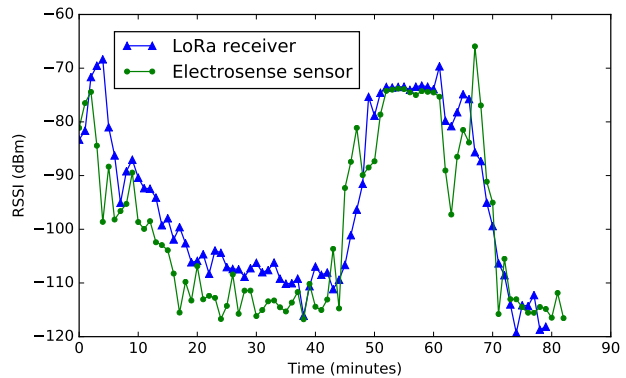

Fig. 6: RSSI measured using a LoRa and Electrosense sensor.

## V. CONCLUSION AND FUTURE WORK

We have introduced Electrosense, a global initiative that seeks a more efficient, safe and reliable use of the electromagnetic space by improving the accessibility of spectrum data for the general public. Electrosense as a crowdsourced network provides solutions to the problems associated with large scale spectrum monitoring and the resulting data deluge. The network users receive incentives in terms of free data storage, readily available applications and an open API.

As the first version of the Electrosense network is fully functional, the next step will be in the direction of enabling usable applications in the backend. The feasibility of various applications were already presented in the section IV of this paper. Research will be continued on the application space, to develop new algorithms for spectrum prediction and anomaly detection. We also plan to implement temporal and spatial interpolation in the backend to produce spectrum geomaps. In addition, in future the PSD pipeline in the sensor will be enhanced with advanced spectrum estimation techniques such as Welch's method. Furthermore, algorithms to extract useful analytics from the spectrum data to overcome data storage constraints will also be explored in future. Detecting fake or forged data from malicious users, by enabling temporal spectrum comparisons among co-located sensors and user sensor ranking, is another active area for future research. Extensive studies will be also done to understand how complex algorithms can be disintegrated for large scale deployment on these cheap sensors. Electrosense can be the first step to democratize the access to the spectrum data to everyone. The age of spectrum data democratization has arrived and it could help to increase the transparency and the knowledge about spectrum usage.